\documentclass[11pt,twoside]{article}

\usepackage{galev06}
\usepackage{epsf}
\usepackage{psfig}
\usepackage{graphicx}
\usepackage{lscape}

\begin{document}
\title{Galaxy and AGN Evolution in the MIR: a combined Spitzer and X-ray view}   
\author{C. Gruppioni$^1$, F. Pozzi$^2$, A. Comastri$^1$, C. Vignali$^2$ and G. Rodighiero$^3$}   
\affil{$^1$ INAF - Osservatorio Astronomico di Bologna, via Ranzani 1, I-40126 Bologna, Italy\\
$^2$ Dipartimento di Astronomia, Universit\`a di Bologna, via Ranzani 1, I-40126 Bologna, Italy\\
$^3$ Dipartimento di Astronomia, Universit\`a di Padova, vicolo dell'Osservatorio 2, I-35122 Padova, Italy
}    

\begin{abstract} 
A proper analysis of the evolution of sources emitting in the Mid-Infrared is
strongly dependent on their broad-band spectral properties (SEDs) at different 
redshifts and luminosities and on a reliable classification allowing to 
disentangle AGN from star-formation activity. 
The diagnostic diagrams based on the optical line ratios are often ambiguous 
and/or misleading not allowing a proper separation of the galaxy/AGN populations.
Thanks to the combination of deep Spitzer and X-rays data a much better census of the 
hidden AGN activity and dust-obscured star-forming galaxies can be obtained,
constraining galaxy and AGN evolutionary models. 
\end{abstract}
 \vspace{-0.2cm}

\section{Introduction}   
Understanding the broad-band Spectral Energy Distributions (SEDs) of galaxies and AGN and their evolutions
is crucial for a complete picture of star-formation and AGN activity in the Universe. At present, only few
local templates are used for modelling the evolution of galaxies and AGN up to high redshifts and the galaxy/AGN 
separation is based mainly on spectroscopy in the optical band, which can be strongly affected by dust absorption.
Thanks to the unprecedented depth reached by {\em Spitzer} in the crucial Mid/Far-Infrared (MIR/FIR) wavelength range
and to the extensive multiwavelength surveys available in specific fields, it is now possible to construct 
broad-band SEDs (from UV to FIR/sub-mm) for large samples of galaxies and AGN up to high redshifts and over a large 
range of luminosities. With the SED analysis it is possible to reveal hidden AGN, escaping optical spectroscopic 
classification, manifesting themselves only in the MIR domain and not dominating at other wavelengths. 
Here we present two particular works aimed at a better census of galaxy and AGN activity through the SED analysis 
of large extragalactic samples in two different areas of the sky: the Hubble Deep Field North (HDFN) 
and the ELAIS-SWIRE field S1 (ES1). 
  
\section{Hidden Activity in Spheroidal Galaxies in the HDFN}  
\label{spheroids}

\begin{figure}

\hspace{1cm}
\includegraphics[height=4.5cm,width=12cm]{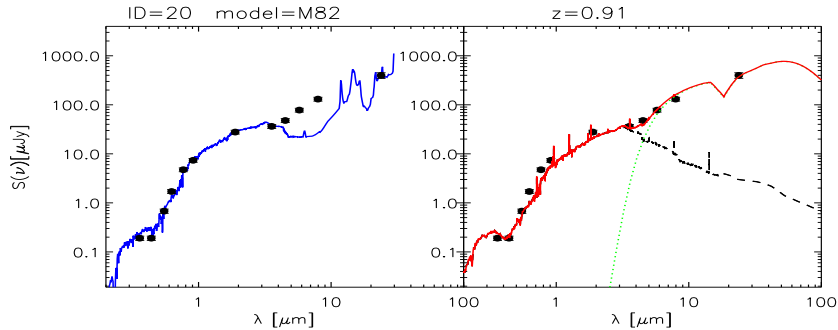}
\caption{Example SED of a spheroidal galaxy with detected 24 $\mu$m
flux. The left panel shows the comparison between the observed SED and a local starburst template spectrum (M82).
The right panel shows a two-component fit (total: solid line) to the same SED: an evolved stellar population (dashed) and 
a dust obscured AGN emission (dotted).}
\label{fig_sed}
\vspace{-0.3cm}
\end{figure}

In the first work (see Rodighiero et al. 2006) we exploit very deep MIR and X-ray observations 
by $Spitzer$ and $Chandra$ 
in the HDFN to identify signs of hidden AGN activity in spheroidal 
galaxies between $z\simeq 0.3$ and 1. Our reference is a complete sample of 168 morphologically 
classified spheroidal (elliptical/lenticular) galaxies with $z_{AB}<22.5$, selected from GOODS 
ACS imaging (Bundy et al. 2005).
Nineteen of these have 24-$\mu$m detection in the GOODs catalogue, 65\% of which 
have an X-ray counterpart in the $Chandra$ catalogue (Alexander et al. 2003).
The nature of the observed MIR emission is investigated by modelling their SEDs based on the 
available multi-wavelength photometry (X-ray, UV, optical, NIR, MIR and radio) and optical spectroscopy. 
The left panel in figure \ref{fig_sed} provides an example of the comparison of model 
fits based on local 
templates (including a variety of spectra of early-type, late-type and starburst 
galaxies, ULIRGs, obscured and unobscured AGN, from Polletta et al.in preparation) with the 
observational SEDs for the early-type sample galaxies. 
In many cases the observed 24 $\mu$m fluxes show a significant excess compared 
with the expectations based on the local spectral templates matched to the optical-NIR data.
In all cases, no acceptable fits with elliptical templates are found, the amount of dust derived from the 
IR emission appearing in excess of that expected by mass loss from evolved stars.
The right panel in Fig. \ref{fig_sed} illustrates a two component (old population $+$ 
dusty torus) solution, provides excellent fits to the data in most cases.

We have used in addition a variety of diagnostics, including the X-ray properties, to interprete the nature of
the energy source in these galaxies. Our multi-wavelength analysis leads us to 
conclude that about half of the IR-detected sources should hide an obscured AGN,
while the X-ray undetected sources are more likely dominated by ongoing star formation.
Therefore, about 15\% -- 20\% of the original sample of 168 spheroidal galaxies in 
HDFN are detected during phases of prominent activity (either AGN or starburst).

\section{SED Analysis of a Complete MIR Extragalactic Sample}
\label{s1_15mic}

In the second work (Gruppioni et al. in preparation) we try to quantify the AGN activity in
a complete sample of MIR selected sources, of which we analyse the broad-band SEDs.
Our reference sample consists of 200 extragalactic sources selected at 15 $\mu$m in the
ES1 field (Lari et al. 2001) with optical spectroscopy and measured redshift (La Franca et al. 2004). 
These sources, for which an extensive multi-wavelength data-set
is available (from X-rays to radio), have previously been considered for obtaining the first derivation of 
the galaxy and AGN Luminosity Function (LF) at 15 $\mu$m (Pozzi et al. 2004; Matute et al. 2006).
We have interpreted the observed SEDs of our MIR sources [from far-UV ($GALEX$-DIS) 
to FIR ($Spitzer$-SWIRE), including optical (B, V, R),
NIR (J, Ks) and MIR/FIR (from 3.6 $\mu$m to 160 $\mu$m) data] by fitting 
them with a library of 21 template SEDs of local objects, the same as used in the
analysis described in section \ref{spheroids} (Polletta et al. in preparation).
We first notice a clear trend from early-type (S0, Sa) to later-type (Sd, starburst) SEDs
with increasing 15 $\mu$m luminosity, while the type-2 AGN are spreaded over the entire range
of L$_{15 \mu m}$ ($\sim$10$^8$ -- $10^{11}$ L$_{\odot}$) and the type 1's show the higher
luminosities ($> 10^{10}$ L$_{\odot}$).

\begin{figure}
\vspace{-0.2cm}

\hspace{2cm}
\includegraphics[height=5cm,width=8cm]{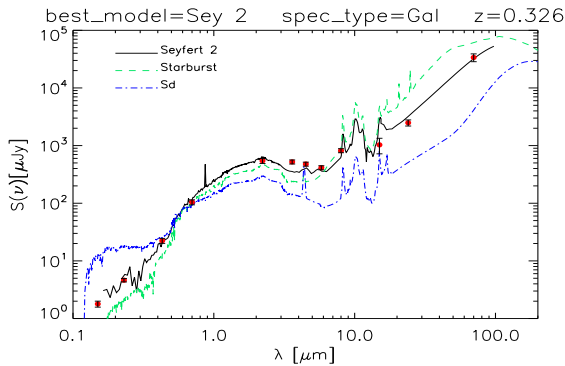}
\caption{Observed SED (filled circles with error-bars) of a MIR source spectroscopically classified as galaxy,
but best-fitted by a Seyfert 2 template (solid line). For comparison, also the template SEDs of a starburst
galaxy (M82, dashed line) and of a Sd galaxy (dot-dashed line) are shown, normalized to the visual band value.} 
\label{fig2}
\end{figure}
The main result comes from the comparison of the SED classification with the spectroscopic 
classification: we find, in fact, that a significant number of previously classified galaxies 
on the basis of their optical spectra are now best-fitted with a Seyfert 2 template SED. An example is
shown in Fig. \ref{fig2}, where for comparison the template SEDs of a starburst galaxy and of a Sd
galaxy are shown together with the observed data points and the best-fit template SED (Seyfert 2). 
From the plot is also clear that the range of wavelengths covered
by IRAC (3.6 -- 8 $\mu$m) is crucial for disentangling AGN from star-formation activity: in fact,
is just in this range that some sources show a "flat" spectrum, incompatible with 
star-forming galaxy template SEDs (from S0 to extreme starbursts). 
\begin{figure}

\plottwo{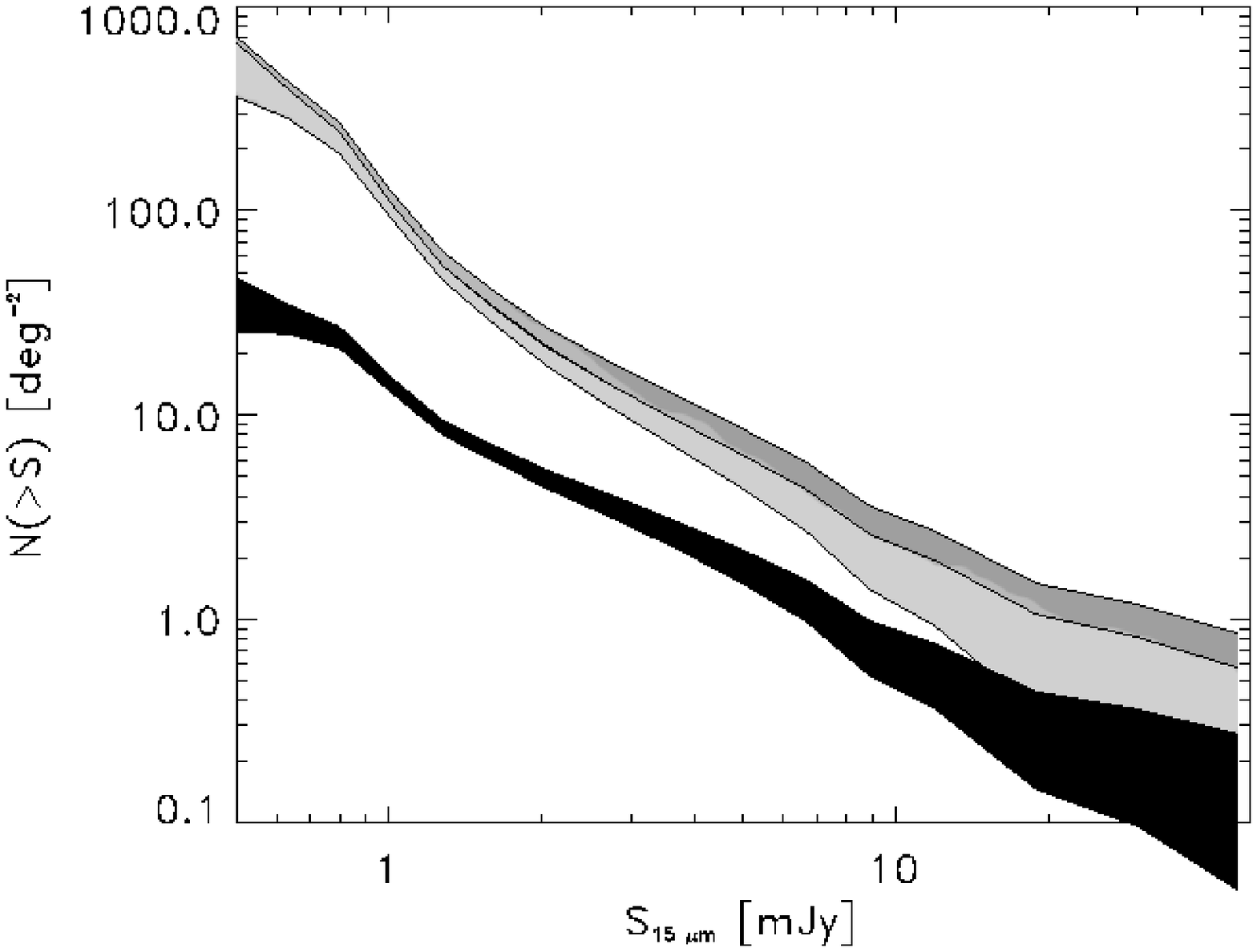}{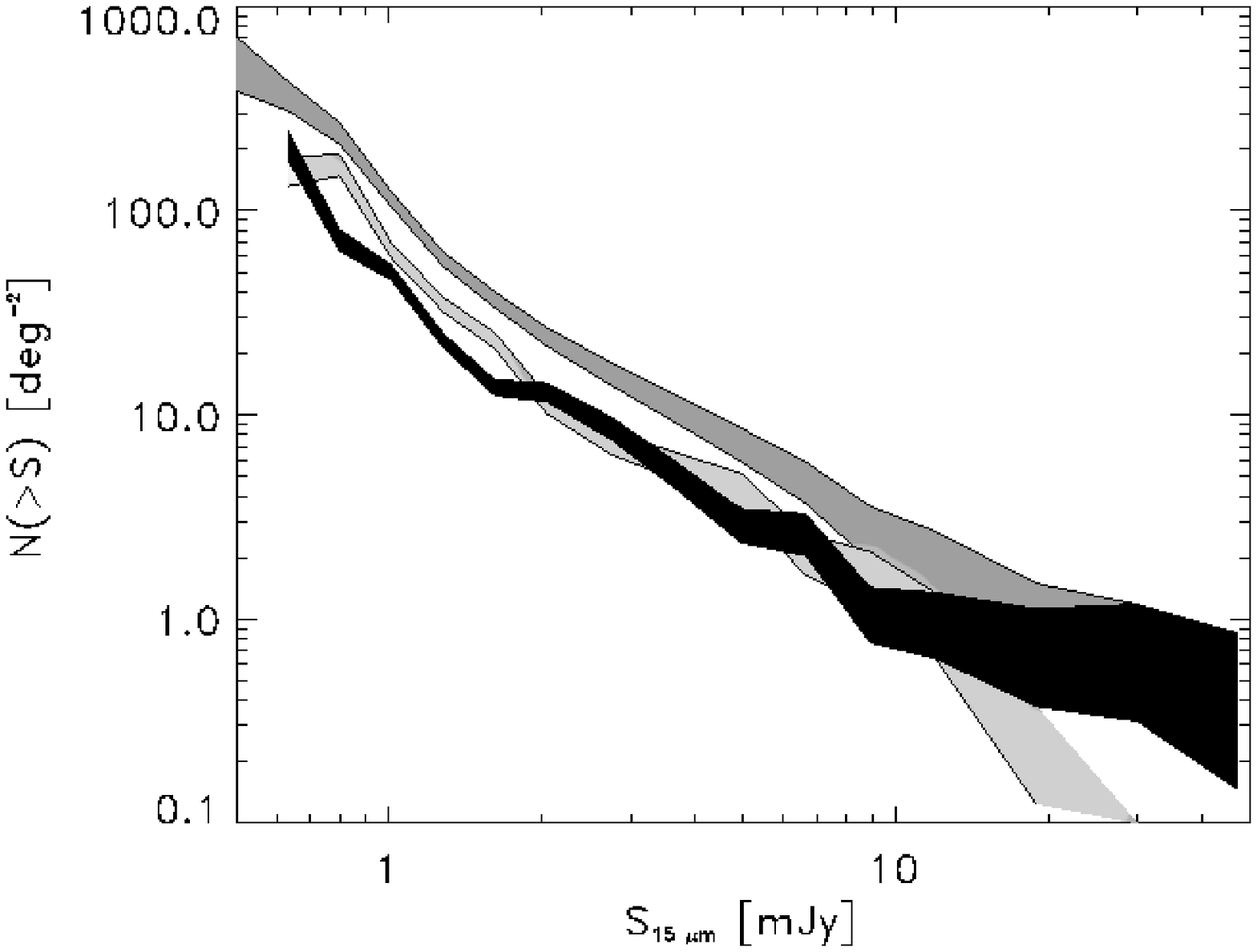}
\vspace{-0.5cm}

\caption{Observed 15 $\mu$m extragalactic source counts in the ELAIS-SWIRE field S1 (total: dark grey area) and 
relative fractions of galaxies (light grey) and AGN (black) as derived from spectroscopy (left panel) and from
the SED-fitting analysis (right panel).} 
\label{fig3}
\end{figure}
The spectroscopic type 1 AGN do show SEDs
in agreement with the spectral classification, while the spectroscopic type 2 AGN are
now fitted in about the same proportions by Seyfert 2's, Power-law AGN's and star-forming galaxy's 
template SEDs. However, the over-all trend is that, thanks to the crucial {\em Spitzer} data,
the SED analysis allows to identify a larger fraction of AGN among MIR sources than optical 
spectroscopy. This is probably due to the fact that these "obscured" AGN do not dominate in 
the optical-UV range, where their spectra show lines from the host galaxy, while the dusty 
torus shows up in the MIR, producing a flat SED from $\sim$3 to 8 $\mu$m, too flat to be 
fitted by galaxy templates. Although the results of our work might be affected by photometric errors in the 
data, not taken into account in our analysis, and by some degrees of degeneracy in the template SEDs, we 
are pretty confident that the majority of the SED-classified AGN are reliable.
However, due to the possible sources of uncertinty, we can consider the fraction of AGN found by the SED-fitting 
analysis as an upper limit to the AGN fraction in MIR selected samples.
Therefore, we can compare the upper-limit to the AGN fraction as derived by our SED-fitting 
analysis to the lower-limit coming from the spectroscopic classification (and considered
for deriving the AGN LF at 15 $\mu$m by Matute et al. 2006) and determine how the relative 
source counts of galaxies and AGN should change according to the new values. The extragalactic
source counts at 15 $\mu$m in the ES1 field (Gruppioni et al. 2002) with the relative contribution of 
galaxies and AGN computed according to the
previous and new determination of AGN fractions are shown in the left and right panels of Fig. \ref{fig3}
respectively.    
In the light of the new results obtained with the SED-fitting analysis, the existing models of galaxy and AGN
evolution in the IR must be revised taking into account the higher fraction of AGN dominating in the MIR wave-range
(this will be addressed in a future paper by Gruppioni et al. in preparation).



\end{document}